\documentclass{INTERSPEECH2023}

\interspeechcameraready 
\usepackage{xcolor}
\usepackage{subcaption}
\usepackage{multirow}

\title{Preference-based training framework for automatic speech quality assessment using deep neural network}
\name{Cheng-Hung Hu$^1$, Yusuke Yasuda$^1$, Tomoki Toda$^1$}
\address{
  $^1$Nagoya University}
\email{hu.chenghung@g.sp.m.is.nagoya-u.ac.jp, yasuda.yusuke@g.sp.m.is.nagoya-u.ac.jp, tomoki@icts.nagoya-u.ac.jp}

\begin{document}

\maketitle
 
\begin{abstract}
One objective of Speech Quality Assessment (SQA) is to estimate the ranks of synthetic speech systems. However, recent SQA models are typically trained using low-precision direct scores such as mean opinion scores (MOS) as the training objective, which is not straightforward to estimate ranking. Although it is effective for predicting quality scores of individual sentences, this approach does not account for speech and system preferences when ranking multiple systems. We propose a training framework of SQA models that can be trained with only preference scores derived from pairs of MOS to improve ranking prediction. Our experiment reveals conditions where our framework works the best in terms of pair generation, aggregation functions to derive system score from utterance preferences, and threshold functions to determine preference from a pair of MOS. Our results demonstrate that our proposed method significantly outperforms the baseline model in Spearman's Rank Correlation Coefficient.

\end{abstract}
\noindent\textbf{Index Terms}: Speech Naturalness Assessment, Speech Quality Assessment, Pairwise Comparison, MOS

\section{Introduction}
\label{sec:introduction}
Speech quality is usually used as a criterion assessing the performance of speech applications such as hearing aids \cite{hines2012speech}, VoIP \cite{hines2015visqol}, speech synthesis systems \cite{glowtts,gradtts}, speech coding systems \cite{garbacea2019low, lpcnet}, etc.
To determine the speech quality of a system, subjective evaluation methods like ITUT Recommendation P.85 \cite{P85} are commonly used. 
However, it is resource-exhausting and time-consuming to collect a listener-unbiased result.
Therefore, it is essential to develop an automatic and reliable method for speech quality assessment (SQA).

Recently, several data-driven SQA approaches are proposed using deep neural network (DNN) \cite{MOSNet, mbnet, ldnet, sqapp, UTMOS, noresqamos} to learn Mean Opinion Score (MOS) of utterances. 
When training models with subjective scores like MOS, there are two potential issues to consider. 
One problem, as noted by Manocha et al. \cite{noresqa}, is the lack of references. 
This can be a challenging problem to address since the model are expected to learn the implicit distribution of references used by human listeners, whether consciously or unconsciously. 
This reference distribution can be heavily influenced by the listener's mood or experience. 
Despite the efforts of several SQA models to learn the distribution using listener IDs to identify each individual for each MOS \cite{mbnet, ldnet}, the lack of reference distribution remains a significant challenge. 
Second, the size of the test set is also a problem. When the number of sentences from a system in the test set is particularly small, determining the system quality score by a small number of utterance scores can result in significant noise.

A preference score is another subjective score to assess speech quality. The preference scores are recognized to be easier and faster to evaluate for human raters than direct scores such as MOS \cite{thurstone_model}, which characteristics make preference scores less noisy \cite{DBLP:journals/corr/ShahBBPRW14, teodorescu2016absolutely}. The preference score can also be converted to system quality scores by aggregation methods \cite{thurstone_model,btl}.

In this paper, we develop a method to convert MOS from a pair of utterances into the form of preferences, which we hypothesize is more suitable for training SQA models. 
Although we use the derived preference scores instead of real preference scores, this method can still address the aforementioned issues: 1) By explicitly providing a reference for the model, the model no longer needs to learn the distribution of the reference on its own. 2) we propose to use MOS from the same listener to generate preference scores, which can more effectively reduce the listener bias in preference scores, and 3) our method can increase the number of evaluations for each system in the test set, reducing the noise in predicted system quality scores.

Our proposal comprises not only a training framework for SQA models that relies on the derived preference scores as a training objective, but also includes speech pair generation methods, aggregation functions to obtain a system score from utterance scores, and threshold functions to determine the preference from a pair of quality scores. Since the method used to aggregate preference scores differs from that used for utterance scores, although the system quality score derived from preference scores has its own meaning, it may not be linearly correlated with the system quality score aggregated from MOS. Consequently, this paper will focus on evaluating the correlation of system ranks using Spearman's Rank Correlation Coefficient (SRCC), rather than system quality scores using the Linear Correlation Coefficient (LCC).

A preliminary simulation is conducted to show the feasibility of converting MOS to preference scores for training. Further, in the simulation, we also found that using MOS from the same listener can improve the performance bound.
We then conduct our experiments by training the baseline model and our preference score based model. The experimental results showed that our model had a statistically significantly better performance (p-value $<$ 0.05) than the baseline in terms of SRCC.

\section{Proposed method}
\label{sec:proposed_method}
\subsection{Framework}
\begin{figure}[t]
  \centering
  \begin{subfigure}[b]{0.5\textwidth}
    \includegraphics[width=\textwidth]{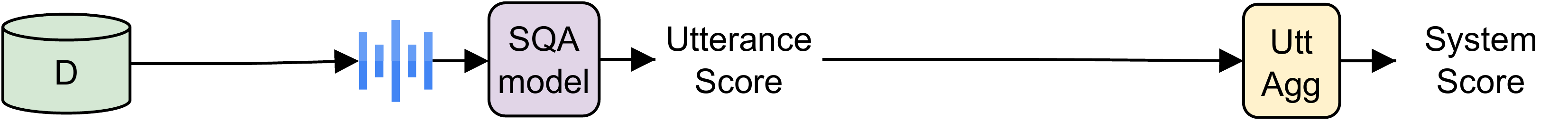}
    \caption{General Non-Reference Speech Quality Assessment model.}
  \vspace*{3mm}
    \label{fig:NRSQA}
  \end{subfigure}
  \begin{subfigure}[b]{0.5\textwidth}
    \includegraphics[width=\textwidth]{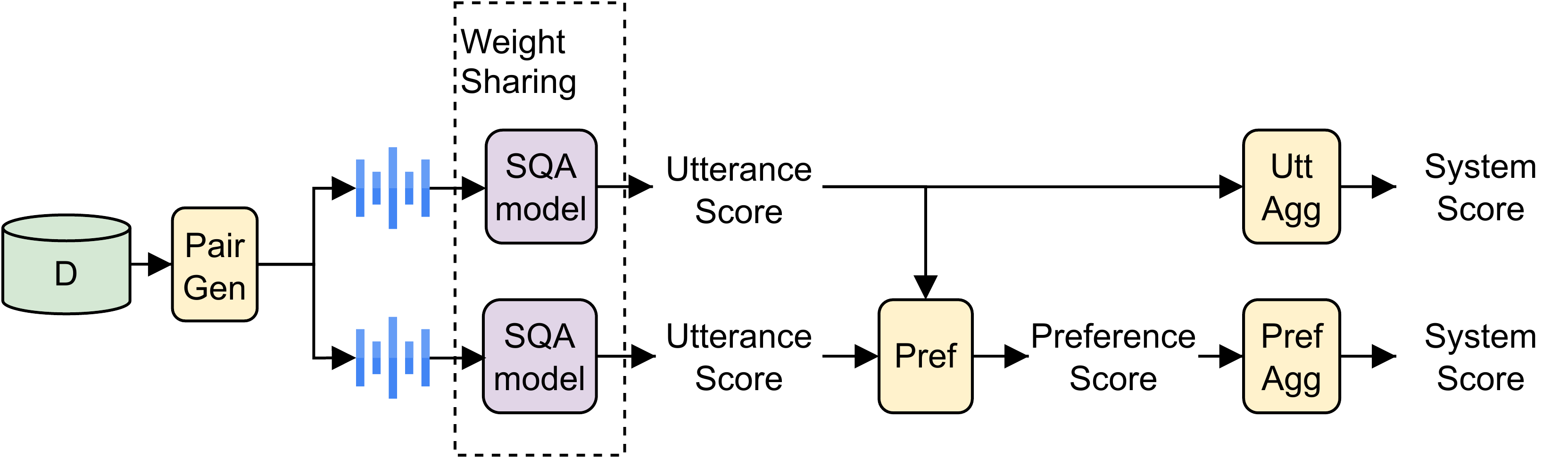}
    \caption{Our proposed model. D, Pair Gen, Pref, Utt Agg, Pref Agg denote the utterance dataset, Pair Generation, Preference Function, Utterance Aggregation Function, and Preferential Aggregation Function, respectively. The same SQA model is used for predicting utterance scores from utterances.}
    \label{fig:proposed}
  \end{subfigure}
  \vspace{-6mm}
  \caption{Framework of speech quality assessment models.}
  \label{fig:models}
  \vspace{-6mm}
\end{figure}

Figure~\ref{fig:models} shows a framework of our proposed method along with a normal general-non-reference SQA method. As shown in Fig.~\ref{fig:models} (a), a normal SQA method predicts two scores in different levels: (1) an utterance score predicted for a single utterance, and (2) a system score derived by aggregating all utterance scores based on an aggregation function. Our method shown in Fig.~\ref{fig:models} (b) predicts a preference score in addition to the utterance score and the system score: (1) a preference score predicted from a pair of the utterance scores based on a preference function, and (2) a preferential system score based on a preferential aggregation function. This framework enables our SQA model to be trained with comparative scores while predicting quality scores in the same way as general-non-reference SQA models.
\vspace{-1mm}
\subsection{Pair Generation}
\label{sec:pair_generation}
In the training phase, we randomly select a listener and then choose two utterances that are assessed by the listener. Note that the utterance pair can contain different content.
In the testing phase, a subset of all possible combinations of utterance pairs are generated by proposed pair generation methods.
Given $N$ systems and $K$ pairs to generate, we consider:

\noindent \textbf{Random Selection (RAND).} We randomly select a pair of systems $\mathrm{(sys_i, sys_j)}$ from all possible system pairs and increment a counter for that system pair. The process is repeated until a total of $K$ pairs are generated. In this method, each system may be compared a different number of times, and the number of pairs formed between each system may also vary.

\noindent \textbf{Linked Selection (LINK).} 
This method generates an equal number of comparisons for each system.
We perform the following steps K/N times. First, 
We create a circular list of integers $\mathrm{[1, 2, ..., N]}$. Then, we randomly shuffle the circular list to obtain a permutation $\mathrm{[a_1, a_2, ..., a_N]}$ of the integers. Next, we form system pairs by pairing up the consecutive integers in the permutation, as follows: $\mathrm{(sys_{a_1}, sys_{a_2})}$, $\mathrm{(sys_{a_2}, sys_{a_3})}$, ..., $\mathrm{(sys_{a_N}, sys_{a_{1}})}$. Finally, we count the number of occurrences for each system pair.

\noindent \textbf{Balanced System Pair Selection (BS).} 
This method generates all combinations of systems to a total of $K$ pairs.


\vspace{-1mm}
\subsection{SQA Model}
Our model is based on the neural network (NN) part of UTMOS \cite{UTMOS}, which is the state-of-the-art model for the VoiceMOS dataset \cite{voicemos}.
The NN of UTMOS has five inputs: the data-domain ID, the listener ID, the phoneme sequence, the reference sequence, and the SSL feature.
Phoneme sequence is recognized by pretrained ASR \cite{asr_utmos} model and then clustered by DBSCAN algorithm \cite{dbscan} to generate the reference sequence.
The SSL feature is extracted from the pretrained wav2vec2 \cite{wav2vec2} model.
These inputs are concatenated and fed to the subsequent BLSTM layer and linear layers to produce the frame-wise scores. The frame-wise scores are then averaged to form the utterance score. The loss of the original UTMOS is calculated by the contrastive loss and the clipped MSE loss.
\subsection{Preference Function}
We derive a preference score from a pair of utterance scores with a preference function. The preference function is used in the bottom path in Fig. \ref{fig:models} (b).
%

Given the $a$-th subjective listening test result of system $i$, an utterance $x_{i,a}$ is assessed by a listener $l_{i,a}$,
the predicted preference score $\mathrm{pref_{pred}}(i,a,j,b)$ is calculated as:
\begin{equation}
  \mathrm{pref_{pred}}(i,a,j,b) = \alpha(\mathrm{SQA}(x_{i,a}, l_{i,a}) - \mathrm{SQA}(x_{j,b}, l_{j,b}))\nonumber
  \label{pref_pred}
\end{equation}
, where $\alpha(x) = 2 \mathrm{sigmoid}(x) - 1$ and $SQA(\cdot, \cdot)$ is the output of the SQA model. Note that one listener can assess more than one utterances, that is, there is $l_{i,a} = l_{j,b}$ for some $i$, $a$, $j$, $b$.
The ground-truth preference score is defined as
$\mathrm{pref_{gt}}(i,a,j,b) = \mathrm{sgn}(s_{i,a} - s_{j,b})$,
where $\mathrm{sgn}(\cdot)$ is the sign function, and $s_{i,a}$ is ground-truth MOS of the utterance $x_{i,a}$ assessed by the listener $l_{i,a}$. 
We use Mean Square Error (MSE) 
as a training objective:
$L = \mathrm{MSE}(\mathrm{pref_{pred}}, \mathrm{pref_{gt}})$.

\begin{figure*}[t]
  \begin{subfigure}[b]{0.33\textwidth}
    \includegraphics[width=\textwidth]{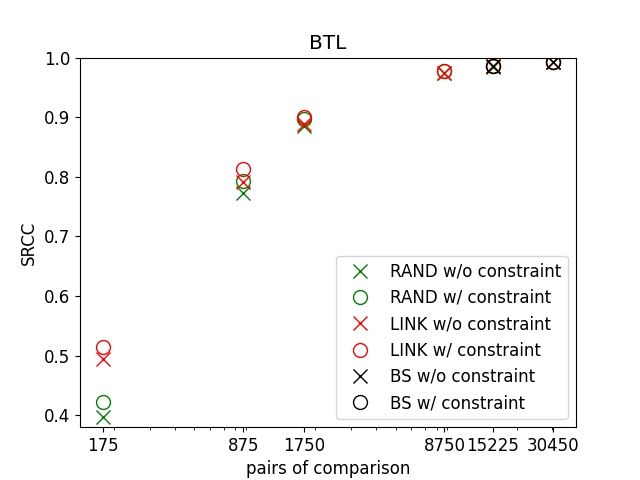}
    \caption{}
    \label{fig:btl_srcc}
  \end{subfigure}
  \begin{subfigure}[b]{0.33\textwidth}
    \includegraphics[width=\textwidth]{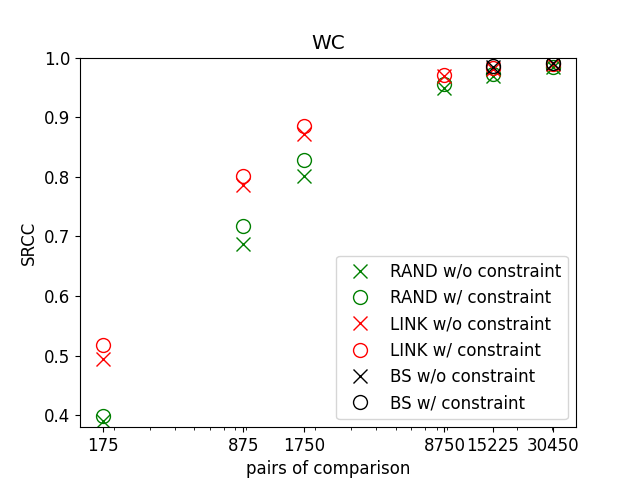}
    \caption{}
    \label{fig:wc_srcc}
  \end{subfigure}
  \begin{subfigure}[b]{0.33\textwidth}
    \includegraphics[width=\textwidth]{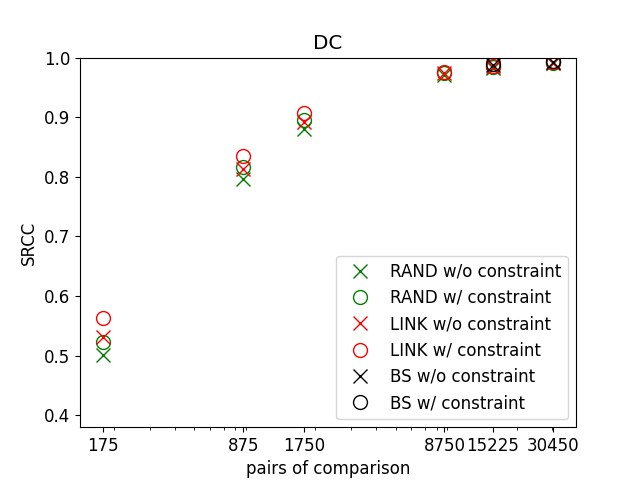}
    \caption{}
    \label{fig:dc_srcc}
  \end{subfigure}
  \vspace{-4mm}  \caption{Simulation results of (a) BTL model, (b) WC aggregation, (c) DC aggregation against ground-truth MOS.}
  \label{fig:srcc}
  \vspace{-6mm}
\end{figure*}

\subsection{Preferential Aggregation Function}
The normal SQA models use the average function to aggregate utterance scores into system scores.
Our preference-based SQA method uses various preferential aggregation functions along with threshold functions to determine a win, draw, or loss to derive system quality scores from preferences of utterance pairs.


\subsubsection{Threshold selection}
We determine the threshold by three methods:

\noindent \textbf{Equal Range (ER).} 
We split the range 
$[-1, 1]$ of $\mathrm{pref_{pred}}$
into three equal intervals, [-1, -1/3], [-1/3, 1/3], and [1/3, 1] to represent 
a lose, draw, and win, respectively.

\noindent \textbf{Equal Error Rate (EER).} We use the development set to find two equal error rate thresholds. One threshold is found between a win and a non-win, and the other is found between a lose and a non-lose. Empirically, the bound is nearly at 0.15 and -0.15.

\noindent \textbf{No Draw (ND).} 
We ignore the draw condition. $\mathrm{pref_{pred} > 0}$ means a win while $\mathrm{pref_{pred}< 0}$ means a lose.

\subsubsection{Aggregation method}
Reduction from preferences to absolute values is studied in the utility theory \cite{UtilityTheoryForDecisionMaking}. The utility theory associates latent utility values with preference probability. In our case, the latent utility represents the absolute quality of a system. The utility model formulates the preference probability $p(i \succ j)$ of $i$ over $j$ based on a difference of their utility values $u_i, u_j$ and the link function $\sigma$ as $p(i \succ j) = \sigma(u_i - u_j)$. The link function works as cumulative probability distribution to convert the utility difference to preference probability \cite{DBLP:conf/nips/Kumagai17}. We can aggregate preferences into absolute values by obtaining utility values.

We use the following aggregation methods:

\noindent \textbf{Differential Count (DC).} We use the win count minus the lost count as the quality score of a system.
This is equivalent to using a linear function $\sigma(x) = \frac{1 + x}{2}$ as the link function to derive preference probability from the utility difference, where $x$ is the utility difference $x = u_i - u_j$. This setting assumes the distribution of the utility difference is uniform.

\noindent \textbf{Bradley-Terry-Luce (BTL) model.} The Bradley-Terry-Luce (BTL) \cite{btl} model is a well-known probability model for deriving absolute scores from pairwise comparisons. 
The BTL is equivalent to using the sigmoid function $\sigma(x) = \frac{1}{1 + \exp(-x)}$ as the link function to derive preference probability from the utility difference. This setting assumes the distribution of the utility difference is logistic. Note that if exponential utility $q_i$ is used as $u_i = \log q_i$, the preference probability becomes the ratio of the exponential utilities: $p(i \succ j) = \frac{q_i}{q_i + q_j}$.
To obtain the utility values, the BTL model iteratively updates the utility values starting from constants based on preference data.
We set the maximum number for iteration to 200 and the tolerance to 0.0001. 

\noindent \textbf{Wining Count (WC).} We use the winning count
alone
as the quality score of a system.
This is a naive method that can not be associated with preference probability with the utility model.

\noindent \textbf{Preference Score (PS).} 
This method is proposed to aggregate the raw $\mathrm{pref_{pred}}$ into the system quality score without the use of the threshold selection method.
The quality score of the system $i$ can be obtained by computing the sum of $\mathrm{pref_{pred}(i,a,j,b)}$ over all evaluated combinations with the compared system $j$ and indexes $a$ and $b$, and then subtracting the sum of $\mathrm{pref_{pred}(k,c,i,d)}$ over all evaluated combinations with the compared system $k$ and indexes $c$ and $d$.

\section{Experimental Evaluation}
\label{sec:exp}
\subsection{Dataset}
The dataset used for experiments was the main track of the VoiceMOS Challenge \cite{voicemos}. 
In the training set, there were 4,973 unique utterances, each of which was evaluated 8 times, resulting in a total of 39,784 utterance-score pairs. The set includes 175 systems, which were each evaluated between 96 to 288 times, and assessed by 288 listeners who each evaluated 126 to 152 utterances.
For the development set, there were 1,066 unique utterances, each was evaluated 8 times, resulting in a total of 8,528 utterance-score pairs. The set contains 181 systems, which were evaluated between 8 to 296 times, and assessed by 296 listeners who each evaluated 16 to 177 utterances.
The test set consisted of 1,066 unique utterances, each assigned an average quality score. There is no utterance overlap in the three sets.
The set contained 187 systems, each with 1 to 38 unique utterances.
In order to obtain ground-truth system scores for the subsequent experiments, we averaged the ground-truth scores of each utterance within each system.
\subsection{Simulation of Pair Generations and Aggregations}
\label{exp:simulation}
We investigated the upper bound of SQA model's performance that can be learned from the training dataset under combinations of utterance pair generation methods (RAND, LINK, and BS) with or without the same listener constraint and system score aggregation methods (DC, BTL, and WC). 
The threshold selection methods were not investigated here because they were about model predictions. 
RAND has no restrictions on the number of system pairs, while LINK requires multiples of the system counts and BS requires multiples of the system combinations.
Therefore, we used 175 systems and generated a varying number of system pairs for each frequency of once, twice, five, ten, and fifty times for the LINK method, while 15,225 system combinations are used to generate a varying number of system pairs for each frequency of once and twice for BS method.
We evaluated the average of Spearman’s Rank Correlation Coefficient (SRCC) from 100 simulations for combinations of the pair generation and the score aggregation methods by using the training set of the VoiceMOS challenge.
We use the training set for simulation rather than the test set for two main reasons. First, the test set doesn't contain quality scores assessed by individual listeners, which makes it impossible to demonstrate the effect of the same listener constraint. Second, our model is primarily trained using individual listener scores, so even if we simulate using the average score from the test set, it may still be difficult to reflect the performance achieved by training with individual scores.

\subsubsection{Performance bound of pair generation methods}
Figure \ref{fig:srcc} shows the result of the simulation.
It was evident that all the pair generation methods could reach the performance bound close to SRCC=1 around 30,000 comparisons for any combinations with all aggregation functions.
Both the LINK and BS methods demonstrated similar performance across all aggregation methods. 
The performance of LINK started at SRCC=0.5 at 175 comparisons and gradually increased to SRCC=0.8, 0.87, 0.97, 0.982, and finally achieved 0.99 at 30,450 comparisons. Similarly, the BS method also achieved SRCC=0.985 at 15,225 comparisons and SRCC=0.99 at 30,450 comparisons. 
The RAND method achieved SRCC=0.984 at 30,450 comparisons, although it exhibited inconsistent performance on the lower number of comparisons.
We, therefore, concluded that all the pair generation methods were feasible to achieve high performance combined with any aggregation functions by using a sufficiently large number of comparisons.
\subsubsection{Effects of the same listener constraint}
We observed that the same listener constraint on pair generation method consistently yielded more correlated system ranks than with the method without it. The performance gain from the constraint was up to 0.03 for a small number of comparisons, and there were slight improvements even for a large number of comparisons. We, therefore, concluded that the same listener constraint on pair generation was effective.

\subsubsection{Comparison among aggregation methods}

\vspace{-1.5mm}
We found that the WC aggregation method performed the worst among the three aggregation methods with any pair generation methods. 
In particular, given the small number of pairs generated by the RAND method, the WC aggregation method could even have 0.1 lower SRCC compared with the DC and BTL aggregation methods.
We interpreted that the WC performed poorly because this method could not be associated with preference probability.
%
Combined with the LINK pair generation, the DC and BTL aggregation methods showed different performances around 8,750 pairs.
If the number of pairs was 8,750 or greater, BTL aggregation method was better than DC aggregation method. 
On the contrary, if the number of pairs was lower than 8,750, the DC aggregation method had a higher performance bound than the BTL.
Given 15,225 and 30,450 pairs generated by the BS pair generation, the DC aggregation method was better than that of the BS pair generation and BTL aggregation methods.
As a result, we concluded that WC was not an appropriate aggregation method, BTL had a good combination with the LINK pair generation method, and DC had a good combination with the BS pair generation method.

\begin{table}[]
\scriptsize 
\begin{tabular}{l|c|l|l}\toprule

\centering
\multirow{2}{*}{Model} & Score & 34,782  & 69,564 \\
~ & Type & pairs & pairs \\ \midrule
\begin{tabular}[c]{@{}l@{}}
UTMOS$_{aug}$\end{tabular} & \multirow{2}{*}{Direct} & \multicolumn{2}{c}{0.927}  \\
UTMOS$_{noaug}$            & ~ & \multicolumn{2}{c}{0.932} \\ \midrule
UTP\_SC          & Direct                                                       & \multicolumn{2}{c}{0.930}           \\ \midrule
UTP\_LINK\_PS   & Preference                                                      & 0.934      & 0.940*     \\
UTP\_BS\_PS     & Preference                                                      & 0.934*      & 0.940*     \\ \midrule
\footnotesize{UTP\_LINK\_ER\_BTL}   & Preference                                                 & 0.930      & 0.932     \\
\footnotesize{UTP\_LINK\_EER\_BTL}  & Preference                                                 & 0.931      & 0.932     \\
\footnotesize{UTP\_LINK\_ND\_BTL}   & Preference                                                 & 0.931      & 0.932    \\ \midrule
UTP\_BS\_ER\_DC      & Preference                                                 & 0.934      & 0.941*     \\
UTP\_BS\_EER\_DC     & Preference                                                 & 0.934*      & 0.941*     \\
UTP\_BS\_ND\_DC      & Preference                                                 & 0.934        & 0.940*     \\  \bottomrule
\end{tabular} 
\caption{The experimental result of the 
ranking prediction with SQA models
. The top row shows the 
number of testing utterance pairs. The asterisk(*) mark represents 
statistical 
significance 
($p=0.05$)
between our proposed models and UTMOS.}
\label{exp}
\vspace{-9.5mm}
\end{table}

\vspace{-1mm}
\subsection{Experiment of MOS prediction}
\vspace{-1.5mm}
We trained the original version of UTMOS and our preference model 20 times with different seeds. Then, we tested our models based on two combinations of pair generation and aggregation functions based on the simulation results in Section \ref{exp:simulation}: the LINK and BTL combination and the BS
and DC combination. Each combination is applied with three threshold methods: ER, EER, and ND. We denoted our models as 
UTP\_X\_Y\_Z
where X meant the pair generation method and Y meant the threshold method and Z meant the aggregation method. We also evaluated these pair generations with the PS aggregation for preference score prediction and denoted them as 
UTP\_X\_PS.
For reference, we also checked the direct score prediction performance with the averaging aggregation of our preference models, 
and we denoted this system as 
UTP\_SC.
Note that UTP\_SC did not need the pair generation method.
We followed the same configurations as UTMOS including hyperparameters and model parameters except for the data augmentation methods. In concrete, we downsampled all waveforms to 16kHz and normalized the subjective quality score into the range [-1, 1]. The Adam \cite{adam} optimizer was used for training with 4,000 steps of warming up to 
15,000 training steps. The batch size was set to 12, 1, and 1 for training, development, and testing, respectively. As for the data augmentation methods, we observed that speaking rate-changing and pitch-shifting caused a 
degradation
for not only our preference models but also the original version of UTMOS. Thus, we included results of UTMOS models trained with the data augmentation method (UTMOS$_\mathrm{aug}$) and without the data augmentation method (UTMOS$_\mathrm{noaug}$) for comparison, and we set UTMOS$_\mathrm{noaug}$ as the baseline. We evaluated the average of SRCC from 20 times of experiments. For every evaluation, we regenerate testing pairs by the pair generation methods, if they were applied. We checked the statistical significance of our proposed methods against UTMOS$_\mathrm{noaug}$ with a pairwise t-test.

Table \ref{exp} shows the results of MOS predictions as the average of SRCC. 
For direct quality prediction using averaging aggregation, the 
UTP\_SC
showed similar performance to the baseline. This indicated our preference-based training framework did not cause much degradation in the direct quality prediction.

Using proposed pair generation methods, UTP\_LINK\_PS and UTP\_BS\_PS achieved higher performance than the baseline in general. This suggested that carefully-designed pair generations were important to predict ranks of synthetic speech accurately. Their high performance was prominent when a larger test set was used. The results suggest that using LINK and BS pairs is effective in accurately evaluating SQA models.



As for preference quality prediction, all the models using the LINK pair generation and the BTL aggregation did not show improvements. These models did not show improvements against a larger test set as well. 
On the other hand, most models using the BS pair generation and the DC aggregation showed improvement compared to the baseline. The improvements from these models were prominent when a larger test set was used. 
The performance difference between the LINK and BTL combination and the BS and DC combination could be explained by the assumption of the aggregation methods. The BTL assumed the utility difference follows a logistic distribution whereas the DC assumed uniform distribution. Thus, the assumption of high noise on preference was important for our preference-based training framework.
The choice of threshold functions did not impact the prediction performance. Among the threshold functions we used, ND had a weak assumption prohibiting no ties, but our result indicated that the weak assumptions such as no ties did not matter in our framework. There are stronger assumptions on preferences such as total order or stochastic transitivity \cite{DBLP:journals/jmlr/BengsBMH21}, and investigation of these assumptions would be our future work.



\vspace{-2mm}
\section{Conclusion}
\vspace{-1.5mm}
\label{sec:conclusion}
In this paper, we proposed a preference-based framework for SQA. Our framework consisted of pair generation, aggregation functions to derive system scores from utterance preferences, and threshold functions to determine preferences from quality scores. 
Our method help SQA models to learn the reference distribution explicitly and reduce the listener bias.
Our simulation confirmed that our pair generation and aggregation functions had high performance bounds, and the constraint on the pair generation to select utterance pairs from the same listener improved the performance bound by reducing listener bias of MOS. 
Our experiment showed that our methods significantly outperformed the UTMOS baseline in terms of SRCC, and the choice of the aggregation function was quite important for our framework to be effective.
In the future, we plan to collect the real preference scores 
and compare their effects with our framework utilizing derived preference scores from MOS.
\vspace{-2mm}
\section{Acknowledgements}
\vspace{-0.5mm}
This work was partly supported by JST CREST Grant Number JPMJCR19A3.

\bibliographystyle{IEEEtran}

\bibliography{mybib}

\end{document}